\documentclass[useAMS,usenatbib]{mn2e}
\usepackage{epsfig}
\usepackage{ifthen}
\usepackage[draft]{hyperref}

\def\gsim{~\rlap{$>$}{\lower 1.0ex\hbox{$\sim$}}}
\def\lsim{~\rlap{$<$}{\lower 1.0ex\hbox{$\sim$}}}

\def\gf{{\sc Galform}}

\newcounter{AGNDone}
\def\AGN{\ifthenelse{\equal{\arabic{AGNDone}}{0}}{active galactic nuclei (AGN) \setcounter{AGNDone}{1}}{AGN}}
\newcounter{CDMDone}
\def\CDM{\ifthenelse{\equal{\arabic{CDMDone}}{0}}{cold dark matter (CDM) \setcounter{CDMDone}{1}}{CDM}}
\newcounter{CMBDone}
\def\CMB{\ifthenelse{\equal{\arabic{CMBDone}}{0}}{cosmic microwave background (CMB) \setcounter{CMBDone}{1}}{CMB}}
\newcounter{IGMDone}
\def\IGM{\ifthenelse{\equal{\arabic{IGMDone}}{0}}{intergalactic medium (IGM) \setcounter{IGMDone}{1}}{IGM}}
\newcounter{IMFDone}
\def\IMF{\ifthenelse{\equal{\arabic{IMFDone}}{0}}{initial mass function (IMF) \setcounter{IMFDone}{1}}{IMF}}
\newcounter{IRDone}
\def\IR{\ifthenelse{\equal{\arabic{IRDone}}{0}}{infrared (IR) \setcounter{IRDone}{1}}{IR}}
\newcounter{ISMDone}
\def\ISM{\ifthenelse{\equal{\arabic{ISMDone}}{0}}{interstellar medium (ISM) 
\setcounter{ISMDone}{1}}{ISM}}
\newcounter{NFWDone}
\def\NFW{\ifthenelse{\equal{\arabic{NFWDone}}{0}}{Navarro-Frenk-White (NFW) \setcounter{NFWDone}{1}}{NFW}}
\newcounter{PAHDone}
\def\PAH{\ifthenelse{\equal{\arabic{PAHDone}}{0}}{polycyclic aromatic hydrocarbon (PAH) \setcounter{PAHDone}{1}}{PAH}}
\newcounter{PCADone}
\def\PCA{\ifthenelse{\equal{\arabic{PCADone}}{0}}{principal components analysis (PCA) \setcounter{PCADone}{1}}{PCA}}
\newcounter{SDSSDone}
\def\SDSS{\ifthenelse{\equal{\arabic{SDSSDone}}{0}}{\href{http://www.sdss.org/}{Sloan Digital Sky Survey} (SDSS) \setcounter{SDSSDone}{1}}{SDSS}}
\newcounter{SEDDone}
\def\SED{\ifthenelse{\equal{\arabic{SEDDone}}{0}}{spectral energy distribution (SED) \setcounter{SEDDone}{1}}{SED}}
\newcounter{SNeDone}
\def\SNe{\ifthenelse{\equal{\arabic{SNeDone}}{0}}{supernovae (SNe) \setcounter{SNeDone}{1}}{SNe}}
\newcounter{TdFDone}
\def\TdF{\ifthenelse{\equal{\arabic{TdFDone}}{0}}{\href{http://www.mso.anu.edu.au/2dFGRS/}{Two-degree Field Galaxy Redshift Survey} (2dFGRS) \setcounter{TdFDone}{1}}{2dFGRS}}
\newcounter{TMASSDone}
\def\TMASS{\ifthenelse{\equal{\arabic{TMASSDone}}{0}}{\href{http://www.ipac.caltech.edu/2mass/}{Two-Micron All Sky Survey} (2MASS) \setcounter{TMASSDone}{1}}{2MASS}}
\newcounter{UVDone}
\def\UV{\ifthenelse{\equal{\arabic{UVDone}}{0}}{ultraviolet (UV) \setcounter{UVDone}{1}}{UV}}
\newcounter{WMAPDone}
\def\WMAP{\ifthenelse{\equal{\arabic{WMAPDone}}{0}}{\href{http://map.gsfc.nasa.gov/}{\emph{Wilkinson Microwave Anisotropy Probe}} (WMAP) \setcounter{WMAPDone}{1}}{WMAP}}

\newcommand{\pcite}[1]{(\citealt{#1})}

\title{Cold Mode Accretion in Galaxy Formation}
\author[Andrew J. Benson \& Richard Bower]{Andrew J. Benson$^1$ and Richard Bower$^2$\\
$^1$Mail Code 350-17, California Institute of Technology, Pasadena, CA~91125, U.S.A. (e-mail: \href{mailto:abenson@caltech.edu}{\tt abenson@caltech.edu})\\
$^2$Institute for Computational Cosmology, University of Durham, Durham, U.K.}

\begin{document}

\maketitle

\begin{abstract}
A generic expectation for gas accreted by high mass haloes is that it is shock heated to the virial temperature of the halo. In low mass haloes, or at high redshift, however, the gas cooling rate is sufficiently rapid that an accretion shock is unlikely to form. Instead, gas can accrete directly into the centre of the halo in a `cold mode' of accretion. Although semi-analytic models have always made a clear distinction between hydrostatic and rapid cooling they have not made a distinction between whether or not an accretion shock forms. Starting from the well-established {\sc Galform} code, we investigate the effect of explicitly accounting for cold mode accretion using the shock stability model of Birnboim \& Dekel. When we modify the code so that there is no effective feedback from galaxy formation, we find that cold mode accretion is the dominant channel for feeding gas into the galaxies at high redshifts. However, this does not translate into a significant difference in the star formation history of the universe compared to the previous code. When effective feedback is included in the model, we find that the the cold mode is much less apparent because of the presence of gas ejected from the galaxy. Thus the 
inclusion of the additional cold mode physics makes little difference to basic results from earlier semi-analytic models which used a simpler treatment of gas accretion. For more sophisticated predictions of its consequences, we require a better understanding of how the cold mode delivers angular momentum to galaxies and how it interacts with outflows.
\end{abstract}

\begin{keywords}
galaxies: general, galaxies: formation, galaxies: evolution
\end{keywords}

\section{Introduction}\label{sec:Intro}

The process of galaxy formation must begin with gas, initially distributed rather smoothly, collapsing to high densities. Furthermore, to sustain ongoing star formation in galaxies requires a continued input of gas over their lifetimes. As such, the question of how galaxies get their gas has received a great deal of attention over the history of galaxy formation studies. While the initial stages of this collapse are purely gravitational (the gas being dragged along by the gravitationally dominant dark matter), after halo formation hydrodynamic forces come into play and further collapse is mitigated by the interplay of gravity, hydrodynamics and cooling processes.

An accretion shock is a generic expectation whenever the gas accretes supersonically as it will do if the halo virial temperature exceeds the temperature of the accreting gas \pcite{binney_physics_1977}. Models of virialization shocks have been presented by several authors \pcite{bertschinger_self-similar_1985,tozzi_evolution_2001,voit_origin_2003,book_role_2010} with the general conclusion that the shock occurs at a radius comparable to (or perhaps slightly larger than) the virial radius. Much debate has occurred over the existence of such shocks --- their existence has often been assumed in analytic models of galaxy formation\footnote{This has always been understood to be an approximation, valid only in specific mass regimes: \protect\cite{white_galaxy_1991}, in discussing the fate of the gaseous component of a halo, note that ``cooling rates may be short enough for the multiphase structure to survive the shocks, and it is then unclear how the dynamics of the gas component should be modeled.''} since \cite{rees_cooling_1977}.

Recent work has examined these issues in greater detail. Motivated by hydrodynamical simulations (\citealt{fardal_cooling_2001}; see also \citealt{kerevs_do_2005,ocvirk_bimodal_2008,kerevs_galaxies_2009}), which show that a significant fraction of gas in galaxies has never been shock heated, \cite{birnboim_virial_2003} developed an analytic treatment of virialization shock stability. The virial shock relies on the presence of a stable atmosphere of post-shock gas to support itself. If cooling times in the post-shock gas are sufficiently short, this atmosphere cools and collapses and can no longer support the shock. For cosmological halos this implies that shocks can only form in halos with mass greater than $10^{11}M_\odot$ for primordial gas (or around $10^{12}M_\odot$ for gas of Solar metallicity). These values are found to depend only weakly on redshift and are in good agreement with the results of hydrodynamical simulations.

As a result, in low mass halos gas tends to accrete ``cold''---never being shock heated to the virial temperature and instead raining into the halo as cold clumps along filaments\footnote{While this picture seems reasonable on theoretical grounds, it as yet has little direct observational support \protect\citep{steidel_structure_2010}.}. If this gas is to make it into the galaxy it must nevertheless lose its energy, either by drag processes in the halo or through a shock close to the galaxy which turns its kinetic energy into thermal energy which is immediately radiated away. Halos which do support shocks are expected to contain a quasi-hydrostatic atmosphere of hot gas. The structure of this atmosphere is determined by the entropy that the gas gains at the accretion shock and that may be later modified by radiative cooling \pcite{voit_origin_2003,mccarthy_modelling_2007}. The transition from cold to hot mode accretion is not sharp---halos able to support a shock still experience some cold mode accretion (the relative contributions of the cold mode decreasing with halo mass).

The consequences of cold vs. hot accretion for the properties of the galaxy forming from such an accretion flow have yet to be fully worked out.  As \cite{croton_many_2006}
have stressed, the absence of an explicit cold mode may not be important since the
cold gas accretion rate in small haloes is limited by the growth of the halo rather than by the
system's cooling time.  
In contrast, \cite{brooks_role_2009} demonstrate in hydrodynamical simulations that cold mode accretion 
does allow accreted gas to reach the galaxy more rapidly, by virtue of the fact that it does not have to cool but instead merely has to free-fall to the centre of the halo (starting with a velocity comparable to the virial velocity). This results in earlier star formation than if all gas were assumed to be initially shock heated to the virial temperature of the halo. It is also clear that the situation needs to be carefully
reassessed in the presence of effective feedback schemes that prevent excessive star formation,
particularly in the high redshift universe. 

In this work we implement a treatment of cold-mode accretion into the \gf\ semi-analytic model of galaxy formation, following the methodology of \cite{birnboim_virial_2003}. This will allow us to assess the importance of cold-mode accretion for cosmological populations of galaxies across a range of redshifts, and to suggest additional studies of the cold mode which might improve our understanding of its role in the process of galaxy formation. An important aspect is that we are able to explore how feedback and reheating
of cold gas moderate the cold mode. We note that \cite{cattaneo_modellinggalaxy_2006} have previously explored a simpler implementation of cold mode accretion in a semi-analytic model of galaxy formation. They parameterized cold mode accretion by defining a critical mass scale (which had some dependence on redshift) above which accretion switched from cold mode to hot mode. While motivated by the results of the \cite{birnboim_virial_2003} calculation, this parameterization did not capture the full generality of that work as we will attempt to do herein.

The remainder of this paper is arranged as follows. In \S\ref{sec:Model} we describe our implementation of cold mode accretion in the \gf\ semi-analytic model. In \S\ref{sec:Results} we present our results and, finally, in \S\ref{sec:DiscConc} we discuss their implications.

\section{Cold Mode Accretion Model}\label{sec:Model}

We implement cold mode accretion in our semi-analytic model using the results of \cite{birnboim_virial_2003}. \cite{birnboim_virial_2003} derive a simple criterion for virial shock stability which states that
\begin{equation}
 \epsilon_{\rm s}< \epsilon_{\rm s,crit},~~\hbox{where}~~\epsilon_{\rm s} = \rho_0 r_{\rm s} \Lambda(T_1)/u_0^3,
\end{equation}
where $\rho_0$ is the pre-shock gas density, $r_{\rm s}$ is the shock radius, $\Lambda(T)$ is the usual cooling function, $u_0$ is the pre-shock inflow velocity of the gas and $T_1=(3/16)\mu m_{rm H} u_0^2/k_{\rm B}$ is the post-shock gas temperature\footnote{The left hand side of this expression is equivalent, to order of magnitude, to the ratio of sound-crossing and cooling times in the post-shock gas. This provides some physical insight into this condition: if the post-shock gas can cool too quickly sound waves cannot communicate across the halo and thereby form a hydrostatic atmosphere which can support a shock front).}. The stability parameter $\epsilon_{\rm s}$ was found to be $0.0126$ by \cite{birnboim_virial_2003} for a gas with adiabatic index $\gamma=5/3$.
At each timestep of our calculation we compute this criterion for each dark matter halo using the pre-virialization model described by \cite{birnboim_virial_2003} in their \S5.1. We assume that a fraction, $f_{\rm hot}$ of any gas accreted from the \IGM\ during that timestep is added to the usual hot gas reservoir of the halo (which is assumed to be heated to the virial temperature by the virial shock). We adopt a simple model in which
\begin{equation}
 f_{\rm hot} = \left[ 1 + \exp\left( \left\{\epsilon_{\rm s,crit} - \epsilon_{\rm s} \right\}/ \Delta \epsilon \right) \right]^{-1}
\end{equation}
such that $f_{\rm hot}\rightarrow 1$ for $\epsilon_{\rm s}\ll\epsilon_{\rm s,crit}$ and $f_{\rm hot}\rightarrow 0$ for $\epsilon_{\rm s}\gg\epsilon_{\rm s,crit}$ with the width of the transition being controlled by the parameter $\Delta \epsilon$. Cooling and infall of this hot mode gas then follows the standard treatment for our semi-analytic model (see \citealt{benson_galaxy_2010}). The remaining fraction $f_{\rm cold}(\equiv 1-f_{\rm hot})$ is instead added to a new cold mode reservoir in the halo. Gas in this cold mode reservoir is allowed to infall onto the central galaxy of the halo at a rate
\begin{equation}
 \dot{M}_{\rm infall,cold} = \Gamma_{\rm cold mode} M_{\rm cold mode} / t_{\rm ff},
\end{equation}
where $M_{\rm cold mode}$ is the current mass of gas in the cold mode reservoir, $t_{\rm ff}$ is the free-fall time from the halo virial radius to the halo centre and $\Gamma_{\rm cold mode}$ is a parameter of order unity which we will use to adjust the infall rate. Since cold flow gas arrives at the virial radius moving with speed comparable to the virial velocity we may reasonably expect $\Gamma_{\rm cold mode} > 1$. In fact, calculations suggest that cold mode gas reaches the center of the halo in around one fifth of the freefall time (Yuval Birnboim, private communication). The metal and angular momentum content of the cold mode reservoir are also tracked and we assume that metals and angular momentum are transferred from the cold mode reservoir to the central galaxy at rates assuming that the metallicities and specific angular momenta of cold mode reservoir gas elements are all equal.

If a halo merges with a larger halo, its hot and cold mode reservoirs are assigned to the hot and cold reservoirs of the larger halo in just the same way, i.e. the gas from the infalling halo is treated just as if it were accreted from the \IGM. (In practice this process happens gradually as ram pressure forces remove gas from the satellite halo. However, in most cases the transfer of gas from satellite to host happens rapidly after accretion of the satellite.) Our model includes outflows of gas from galaxies, driven by \SNe\ explosions. Any such reheated gas is always added to the hot mode reservoir (after a delay of order a halo dynamical time---see \citealt{benson_galaxy_2010}). The rest of our semi-analytic model remains unchanged and follows the methods described by \cite{benson_galaxy_2010}.

\section{Results}\label{sec:Results}

We find that values of $\epsilon_{\rm s,crit}=0.0126$ (the value found by \citealt{birnboim_virial_2003}), $\Delta \epsilon=0.01$ and $\Gamma_{\rm cold mode}=2.5$ give results which are in reasonable agreement with those found in numerical simulations. It should be noted that these correspond to a very simple model with little fine-tuning---$\epsilon_{\rm s,crit}$ is as predicted by a simple analytic argument, $\Gamma_{\rm cold mode}=2.5$ implies that the cold reservoir is depleted on a fraction of the freefall timescale as expected and $\Delta \epsilon=0.01$ implies a rapid transition from hot to cold mode.

\begin{figure*}
 \begin{tabular}{cc}
 No SNe-driven outflows & Including SNe-driven outflows \\
 \includegraphics[width=75mm,viewport=0mm 10mm 200mm 265mm,clip]{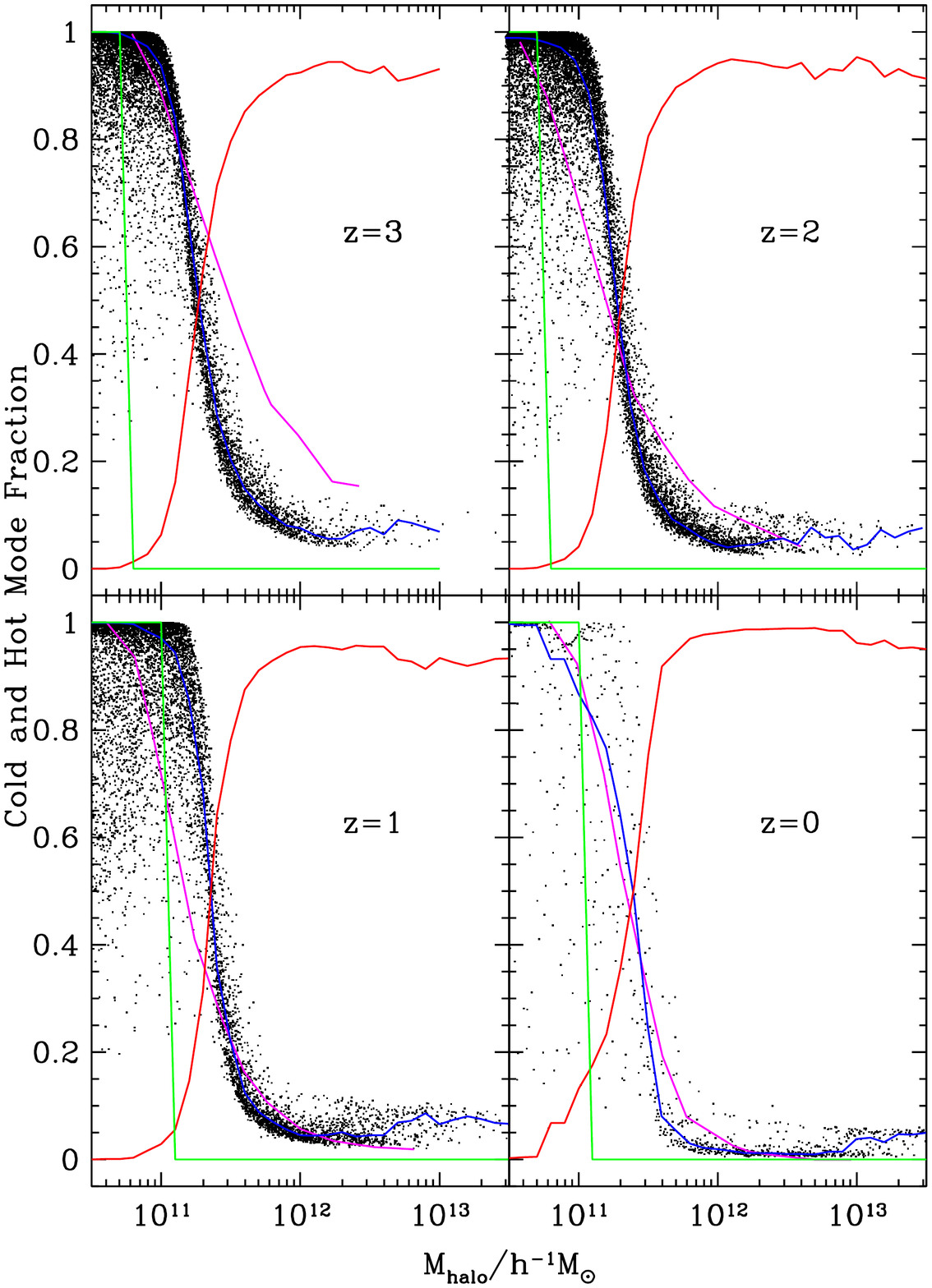} &
 \includegraphics[width=75mm,viewport=0mm 10mm 200mm 265mm,clip]{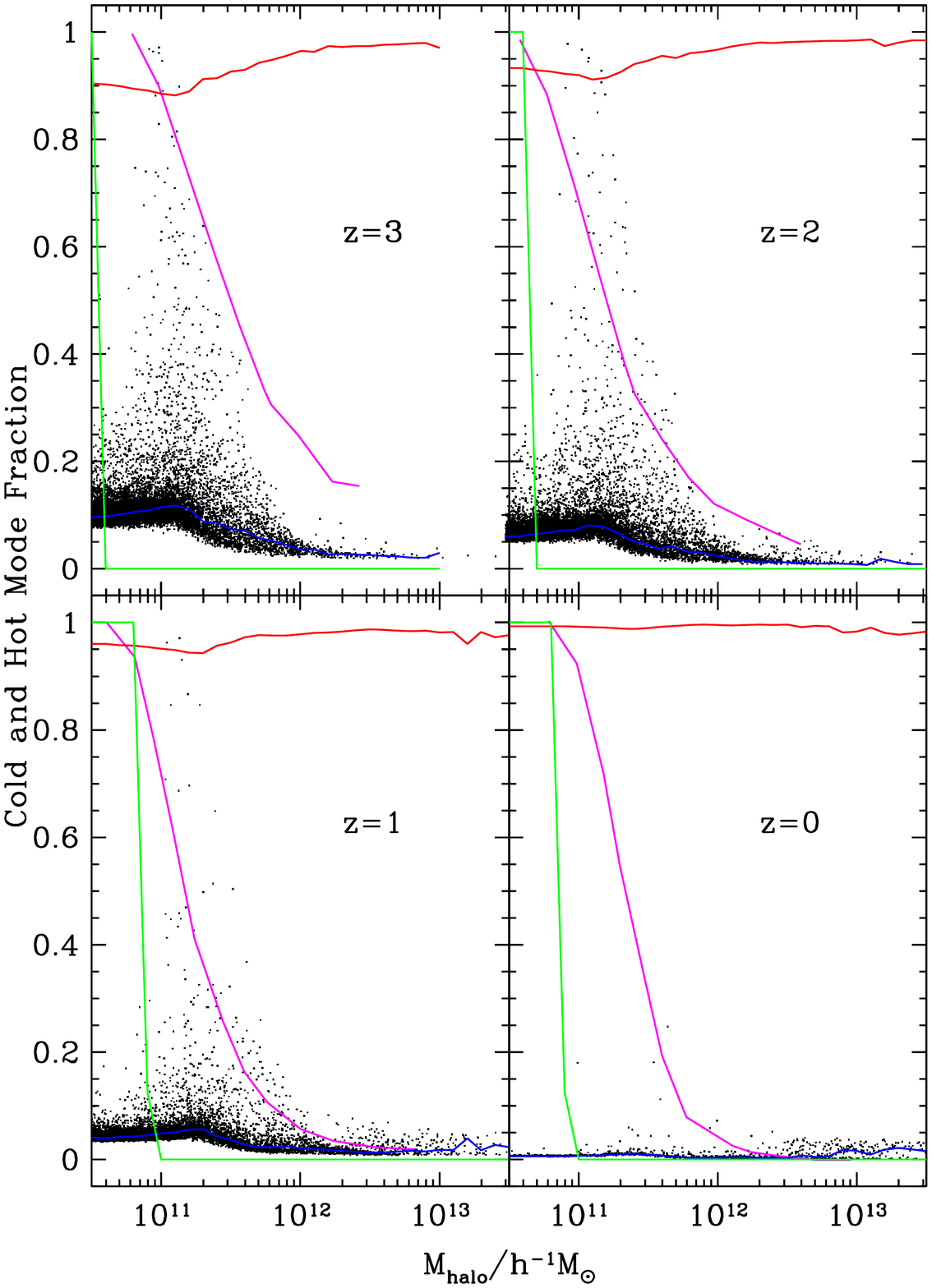}
 \end{tabular}
 \caption{The fraction of diffuse (i.e. non-galactic) gas in halos in the cold mode is shown by the black points as a function of halo mass and for four different redshifts as indicated in each panel. The solid blue line shows the median cold mode fraction at each halo mass. The red line shows the corresponding median hot mode fraction. The magenta lines show the equivalent cold mode fraction from the simulations of \protect\cite{kerevs_galaxies_2009}. Green lines indicate the transition from rapid to slow cooling in models without cold-mode accretion (halos in the rapid cooling regime are assigned a cold fraction of 1, those in the slow cooling regime are assigned a value of 0). \emph{Left-hand panel:} Results when \protect\SNe-driven outflows are not included. \emph{Right-hand panel:} Results when \protect\SNe-driven outflows are included. }
 \label{fig:ColdFracOutflows}
\end{figure*}

In particular, Fig.~\ref{fig:ColdFracOutflows} shows a comparison with the results from hydrodynamical simulations as reported by \cite{kerevs_galaxies_2009}. The fraction of halo gas currently in the cold mode reservoir is indicated by the black points as a function of halo mass and for four different redshifts. The blue lines indicate the median cold mode fraction at each halo mass, while the red lines indicate the corresponding hot mode fraction. For comparison, the magenta lines show the cold mode fraction as reported by \cite{kerevs_galaxies_2009}. For reference, the green lines indicate the mass at which halos transition from rapid to slow cooling (a distinction which has been present in semi-analytic models since their inception). Clearly this occurs at a lower mass scale than the critical mass for shock formation.

In the left-hand panel we show results from our model with \SNe-driven outflows switched off. The simulations of \cite{kerevs_galaxies_2009} do not create strong outflows which may be expected to significantly alter the nature of cold vs. hot mode accretion. Therefore, the fairest comparison to their results is made with outflows removed from our model. Our model produces results in reasonable agreement with those of \cite{kerevs_galaxies_2009}, correctly finding the characteristic transition mass as a function of redshift and producing approximately the correct width of the transition at $z=0$. Our model transitions somewhat too quickly at higher redshifts, which may be indicative of a break-down in the assumption of spherical symmetry in the model of \cite{birnboim_virial_2003}, an assumption which should be more valid at low redshifts. In the right-hand panel of Fig.~\ref{fig:ColdFracOutflows} we switch \SNe-driven outflows back on in our model, under the assumption that gas driven out of galaxies will populate the hot component of the halo. We find that the cold mode fraction is greatly suppressed in this case, since the halos now contain significant quantities of reheated gas. This assumption could be incorrect, with outflowing gas condensing into small, dense clouds; our goal here is merely to indicate how important outflows \emph{could} be in establishing the cold/hot accretion balance in halos.

\begin{figure*}
 \begin{tabular}{cc}
 \includegraphics[width=80mm]{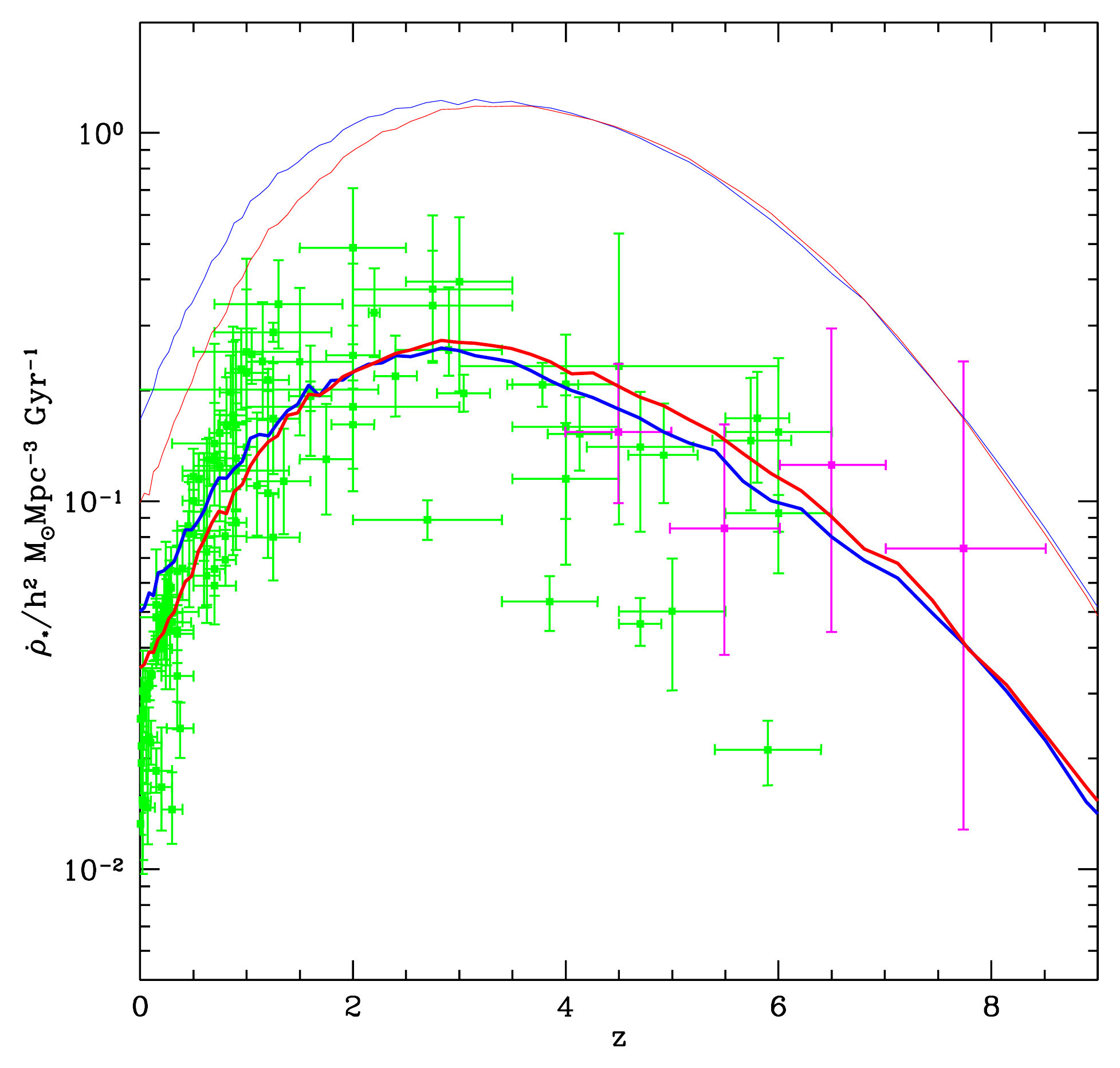} &
 \includegraphics[width=80mm]{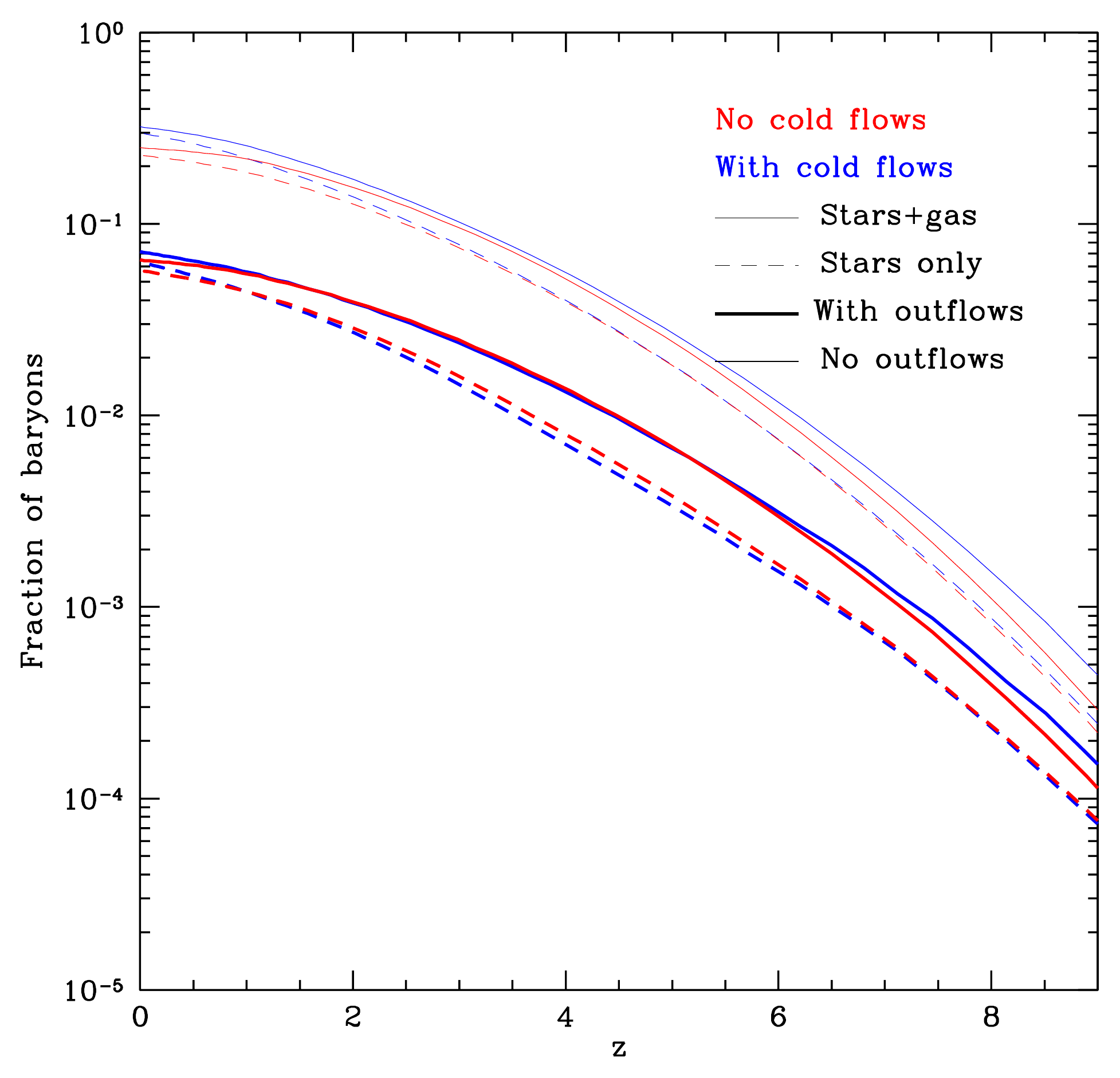}
 \end{tabular}
 \caption{\emph{Left-hand panel:} The volume averaged star formation history as a function of redshift. Green points show observational estimates from a variety of sources as compiled by \protect\cite{hopkins_evolution_2004} while magenta points show the star formation rate inferred from gamma ray bursts by \protect\cite{kistler_star_2009}. Red lines show results for models with no cold-mode accretion, while blue lines are for models including cold-mode accretion. Thick lines indicate models including \protect\SNe-driven outflows, while thin lines are for models without such outflows. \emph{Right-hand panel:} The fraction of baryons locked up into stars (dashed lines) and stars plus galactic gas (solid lines) as a function of redshift. Thick and thin lines correspond to models with and without \protect\SNe-driven outflows. Red lines show models without cold-mode accretion, while blue lines correspond to models that include cold-mode accretion.}
 \label{fig:StarFormation}
\end{figure*}

In Fig.~\ref{fig:StarFormation} we explore the consequences of cold mode accretion for the build up of the galaxy population. The left-hand panel shows the volume averaged star formation rate in models with and without cold mode accretion (blue and red lines respectively) in cases with and without \SNe-driven outflows (thick and thin lines respectively). Of course, the models lacking \SNe-driven outflows produce far too high star formation rates as expected, but are shown for reference. Comparing the blue and red lines, the models have almost identical star formation rates at high redshifts irrespective of the inclusion of cold flows both when \SNe-driven outflows are included and when they are not. At low redshifts, the models including cold mode accretion have higher star formation rates. We note that the model without cold mode accretion and with \SNe-driven outflows has been constrained to fit a variety of observational data, including those shown in this Figure. The addition of cold mode accretion worsens the agreement with the data, but this should not be taken as evidence against cold mode accretion---it is highly probable that adjustments in other parameters could restore good agreement in models with cold mode accretion. Comparing thick and thin lines, the figure emphasizes the importance of using the correct galaxy formation physics to assess the importance of cold flows.  The effects of stellar winds and outflows are far more important (but less certain) than the distinction between cold flows and rapid cooling.

\begin{figure}
 \includegraphics[width=80mm]{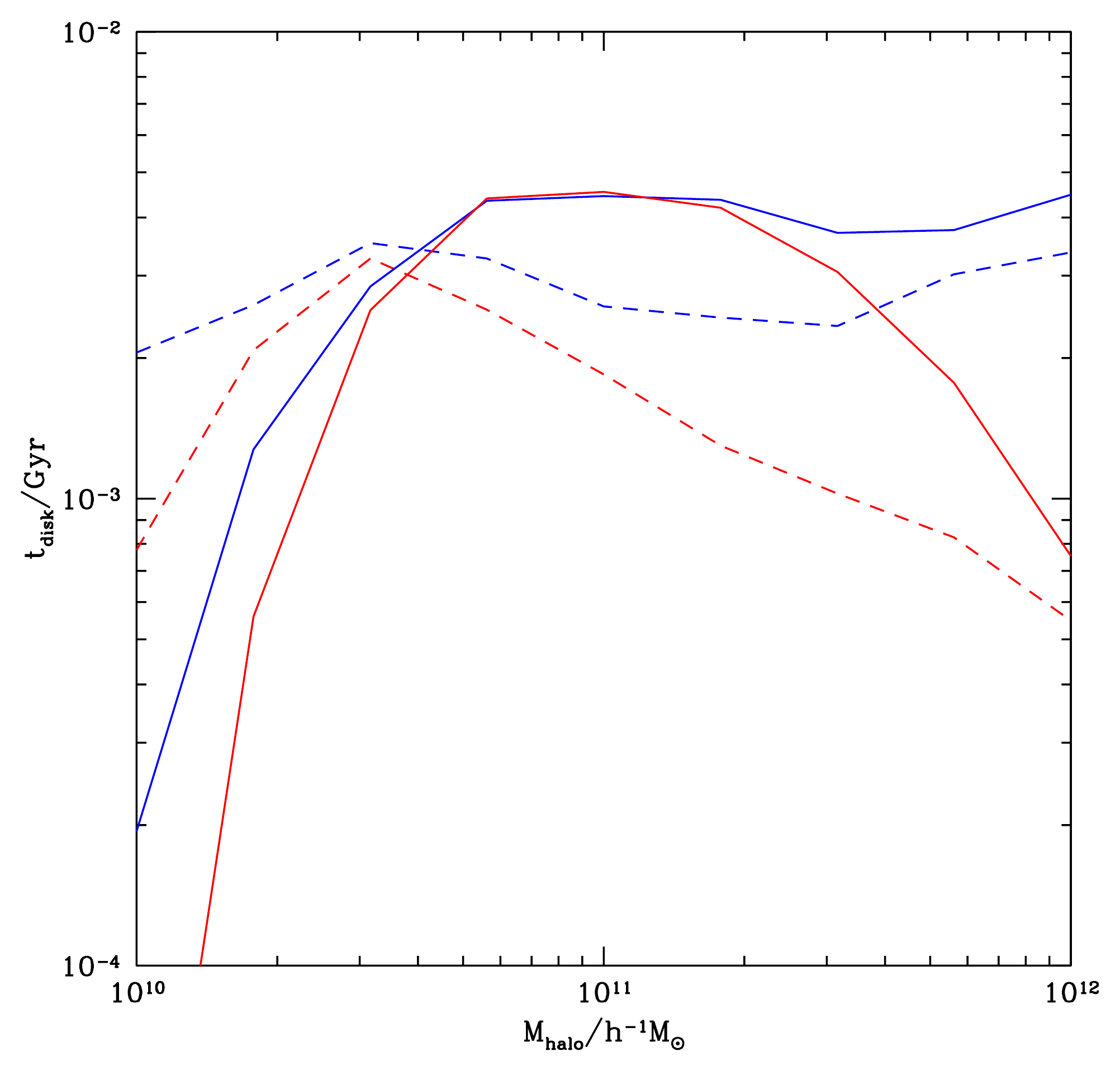}
 \caption{The median dynamical time of galactic disks as a function of halo virial mass at $z=6$. Dashed lines show models with no \protect\SNe-driven outflows while solid lines include \protect\SNe-driven outflows. Red lines show results for models with no cold-mode accretion, while blue lines are for models including cold-mode accretion.}
 \label{fig:tdyn}
\end{figure}

The right-hand panel of Fig.~\ref{fig:StarFormation} shows the fraction of baryons (averaged over the entire universe) in the form of stars (dashed lines) or stars and \ISM\ gas (solid lines) as a function of redshift for the same set of four models. Considering first models with no \SNe-driven outflows, the fraction of baryons in stars is identical in models with and without cold mode accretion at high redshifts as expected from the star formation rate. However, the fraction of baryons in stars plus \ISM\ gas is actually higher in the model including cold mode accretion. This occurs because cold mode accretion is more efficient at getting gas into the galaxy phase but, in our model, results in galaxies with longer star formation timescales due to a higher angular momentum content (and consequently longer disk dynamical times). This is illustrated in Fig.~\ref{fig:tdyn}, where
we show the dynamical times of galactic disks as a function of halo
mass at $z=6$. Dashed lines compare the effect of the cold mode on dynamical
time in the absence of \SNe-driven outflows, while solid lines are for models that include \SNe-driven outflows. In high mass haloes,
the cold mode increases the disk dynamical time by a factor of three or more. In models that include \SNe-driven outflows the effect is weaker
and the difference in total gas content is only significant at higher redshifts. By $z=0$, in models with \SNe-driven outflows, the total mass of baryons in stars and \ISM\ gas differs only slightly between models with and without cold mode accretion.

\section{Discussion and Conclusions}\label{sec:DiscConc}

We have described a simple implementation of cold-mode accretion in a semi-analytic model of galaxy formation, using a previously proposed analytical model with no modifications. Without any adjustment this model provides an excellent match to results from numerical simulations---this could no doubt be improved with some fine-tuning of the model, but our aim here was to demonstrate that the simulation results can be encapsulated by an easy to implement model.

We find that the inclusion of \SNe-driven outflows has a dramatic effect on the fraction of cold mode gas present in dark matter halos---much more so than switching the cold mode accretion on or off --- suggesting that simulations must account for such outflows before they can make robust predictions for the properties of cold mode gas. The inclusion of cold mode accretion makes little difference to luminosity functions and galaxy sizes at $z=0$, implying that results from earlier semi-analytic models which did not treat cold mode accretion are still valid. Without \SNe-driven outflows the cold mode results in a significant increase in the mass and luminosity of brighter galaxies which are able to gain some mass through the cold mode even when their hot mode has been effectively shut down by feedback from \AGN.

Cold mode accretion is found to be effective at getting gas into the galaxy phase at high redshifts, potentially allowing for high rates of star formation during these epochs. However, we find that the star formation rate at $z\gsim 3$ is mostly unaffected when we include cold mode accretion due to the fact that the galaxies that form are larger and lower density than they would be if cold mode accretion were neglected. We caution that the treatment of angular momentum delivery to galaxies via the cold mode is therefore of crucial importance to assessing its impact on star formation rates. To date, this has not been studied in detail in hydrodynamical simulations but should be in order to refine our understanding of how the cold mode affects galaxy formation.

In summary, cold mode accretion should be explicitly accounted for in semi-analytic models --- and will require some retuning of parameters to restore good fits to observational data --- but does not seem to qualitatively change our picture of galaxy formation, at least at the coarse-grained level studied here. More detailed results (luminosity function shapes, distribution of sizes and formation times) will require a significantly more detailed implementation of the cold mode, including how it delivers angular momentum to galaxies (e.g. \citealt{navarro_disk_2009}). Calibration from numerical simulations should improve the accuracy of cold mode results. However, we have seen that the inclusion of cold-mode accretion in to our semi-analytic model does not produce any surprises. The changes in star formation rates and stellar mass fractions, particularly in the low redshift universe are reassuringly small, and much smaller than the differences in galaxy properties created by changes to the feedback effects of galactic winds and AGN.

\section*{Acknowledgements}

AJB acknowledges the support of the Gordon \& Betty Moore Foundation and thanks Alyson Brooks, Brant Robertson and Yuval Birnboim for numerous helpful conversations.

\bibliographystyle{mn2e}
\bibliography{BensonColdMode}

\end{document}